\documentclass[twocolumn, 5p]{elsarticle}

\usepackage{amsmath, amssymb}
\usepackage{breqn}

\usepackage{color}
\usepackage{graphicx}% Include figure files
\usepackage{dcolumn}% Align table columns on decimal point
\usepackage{bm}% bold math
\usepackage[all]{xy}

\newcommand{\beq}{\begin{equation}}
\newcommand{\eeq}{\end{equation}}
\newcommand{\ra}{\rightarrow}
\newcommand{\R}{\mathbb{R}}
\newcommand{\Q}{\mathbb{Q}}

\begin{document}

\title{Stochastic Dynamics of Extended Objects in Driven Systems II: Current Quantization in the Low-Temperature Limit}
%\author {Michael \surname{Catanzaro}$^{a}$}
%\author {Vladimir Y. \surname{Chernyak}$^{a,b}$}
%\author{John R. \surname{Klein}$^{a}$}
\author[mjc]{Michael~J.~Catanzaro}
\author[mjc,vyc]{Vladimir~Y.~Chernyak}
\author[mjc]{John~R.~Klein}

\address[mjc]{Department of Mathematics, Wayne State University,
656 W. Kirby, Detroit, MI 48202}

\address[vyc]{Department of Chemistry, Wayne State University,
5101 Cass Ave, Detroit, MI 48202}

%\affiliation{$^a$Department of Mathematics, Wayne State University,
%656 W. Kirby,Detroit, MI 48202}

%\affiliation{$^b$Department of Chemistry, Wayne State University,
%5101 Cass Ave,Detroit, MI 48202}

%\affiliation{$^c$Center for Nonlinear Studies and Theoretical Division, LANL, Los Alamos, NM 87545}

\date{\today}

\begin{abstract}
Driven Langevin processes have appeared in a variety of fields due to the relevance of natural phenomena having both
deterministic and stochastic effects. The stochastic currents and fluxes in these systems provide a convenient
set of observables to describe their non-equilibrium steady states. Here we consider stochastic motion
of a $(k-1)$-dimensional object, which sweeps out a $k$-dimensional trajectory, and gives rise to 
a higher $k$-dimensional current. By employing the low-temperature (low-noise) limit, we reduce the
problem to a discrete Markov chain model on a CW complex, a topological construction which generalizes
the notion of a graph. This reduction allows the mean fluxes and currents of the process to be expressed in terms
of solutions to the discrete Supersymmetric Fokker-Planck (SFP) equation. Taking the
adiabatic limit, we show that generic driving leads to rational quantization of the generated higher dimensional
current. The latter is achieved by implementing the recently developed tools, coined the higher-dimensional 
Kirchhoff tree and co-tree theorems. This extends the study of motion of extended objects in the
continuous setting performed in the prequel \cite{CCK16b} to this manuscript. 
\end{abstract}

\maketitle

\section{Introduction}
\label{sec:intro}

Non-equilibrium thermodynamics is an area of statistical mechanics dealing
with driven, and often time-varying, systems evolving via chemical reactions and
exchange of energy~\cite{DGM13, J98, HS01, G00, VK92}. These systems
can be accurately described by a Langevin equation when the typical time scale
of the system is slow compared to the relaxation time of the coupled
bath~\cite{LL02, CKW96, R84}. Because of the
generality of this setup, Langevin dynamics have found applications in a 
prodigious array of fields including chemistry~\cite{LL02, G00}, biology~\cite{AA07, A07}, 
finance~\cite{B98,O13, LOT02}, and 
certain engineering applications~\cite{CFBBM15, CCG04, P98}. It is the ubiquitous nature of the Langevin
framework that allows such a pervasive grasp across the physical sciences.

Within the framework of non-equilibrium thermodynamics, the notion of
stochastic current, or flux, has recently gathered interest as a compelling 
observable which gives insight on the underlying system~\cite{SN07, RHJ08, AR14, RHJ09, CCBK13}. Together with 
the probability density, stochastic currents provide an adequate set of observables
for describing the long-time behavior of probability distributions in
this setting. A stochastic flux 
can be measured by counting how many times the system spans a cycle, or equivalently, 
by counting intersections of a stochastic trajectory with a cross section. This
notion of current has been developed completely in the discrete setting on a
mathematical graph, in which case the components of the stochastic current 
are simply given by counting how many times a stochastic trajectory crosses the edge.
In the discrete case of graphs, this Langevin process can be reduced to a 
continuous-time Markov chain. The goal of this manuscript is to extend
the concept of stochastic trajectory to higher dimensions, as well
as the underlying graph on which the trajectory lies.

This manuscript constitutes the sequel to our work on higher-dimensional 
Langevin processes on continuous state spaces of arbitrary dimension. 
In the prequel \cite{CCK16b}, we explained why the proper
way of defining a Langevin process in this setting is in terms of stochastic
vector fields on the underlying state space $X$. Not only does this generalize
the usual definition in a natural way, it allows us to take full advantage of
the dynamical symmetry which is necessary for the reduction to the discrete
model discussed here. After giving several examples of higher dimensional
currents, we defined the generalized Supersymmetric Fokker-Planck (SFP) operator
and equation, which governs the evolution of higher-dimensional objects on $X$.
Using this equation, we were able to write an expression for the average current
generated under a Langevin process. This expression was then re-cast, in the
adiabatic limit, with the solution to the higher dimensional Kirchhoff network
problem and the higher-dimensional Boltzmann distribution. While all of this
was performed in the continuous case, the low-temperature (or low-noise) limit
connects the two manuscripts, as the continuous dynamics tend to a discrete
structure naturally embedded in $X$. In this way, one interested in these
two limits can forget the complicated continuous picture and focus on the 
simpler, more computational discrete setting.

The manuscript is organized as follows. Section~\ref{sec:tech-intro} serves
as a summary of the main ideas and specialized tools used throughout our
calculations. In section~\ref{sec:Markov-chain}, we describe how the low-noise
limit allows for a reduction of the stochastic process to a Markov chain model
defined on the Morse-Smale complex, defined there. 
Section~\ref{sec:generated-currents-CW} contains a formal expression for the
average current generated by a Langevin process. The higher-dimensional
Kirchhoff network theorem is defined and solved in~\ref{sec:Kirchhoff-higher-tree},
whereas the dual higher Boltzmann distribution is given in~\ref{sec:Kirchhoff-higher-cotree}.
These two results are combined in subsection~\ref{sec:low-T-adiabatic-explicit} to
give a much more computable form of the generated current. This version of the current
is used in section~\ref{sec:quantization-CW} to prove the main result of
the manuscript: for generic driving in the low-temperature, adiabatic limit, 
the generated current is rationally quantized.

\section{Technical Introduction}
\label{sec:tech-intro}

In this section, we present the main results and concepts
of the manuscript. Our calculations are technical, and since we
do not assume any topological background, this section serves as
an instructive and brief bridge to our results.

The section is organized as follows. 
Subsection~\ref{sec:CW-Morse-lowtemp} contains a simple and self-contained
presentation of the structures needed for a description of higher-dimensional
discrete stochastic models: CW complex, Morse complex, and CW complex homology.
We also provide an initial insight into how these discrete models, associated with
CW complexes, arise naturally from continuous models in the long-time, low-noise
limit. In subsection~\ref{sec:main results} we present the main
results of the manuscript on fluxes in higher-dimensional discrete models:
higher-dimensional versions of the tree and co-tree Kirchhoff theorems,
explicit expressions for fluxes, and quantization of higher-dimensional fluxes
in the low-temperature adiabatic limit for periodic potential driving.

\subsection{CW complexes, Morse Complex, and Slow Low-Temperature Limit for Stochastic Dynamics}
\label{sec:CW-Morse-lowtemp}

As briefly discussed in section~\ref{sec:intro}, if one is interested in the
low-temperature (low noise) limit on long time scales, stochastic
dynamics on continuous spaces can be reduced to its discrete counterpart. In
the simplest case of a potential force (often referred to as a gradient
flow), the discrete setting starts with a graph whose vertices are represented
by the local minima of the potential $V(x)$, whereas the edges correspond to
saddle points, or more rigorously to critical points of the potential (defined
by the conditions $dV = 0$) with exactly one unstable mode. Equivalently, these are
critical points with {\it Morse index} (which is the number of unstable modes)
equal to $1$. On the initial time scales of low-noise stochastic dynamics, a particle
approaches a local minimum, completely determined by the
deterministic component $u(x)$ of the velocity field in the Langevin equation
(Eq.~(1) of \cite{CCK16b}). After that, the particle will spend most of the
time in a small neighborhood of the local minimum, participating in rare
transitions. These occur when a highly improbable configuration of the noise field takes
the particle to a saddle point, and actually a little bit past it, so that it
further drops to a different local minimum with no additional noise assistance needed.
The rate $k_{a\alpha} \sim e^{-\kappa^{-1}(V(x_{\alpha}) - V(x_a))}$ of a
transition over the edge $\alpha$ that starts at vertex $a$ can be calculated
analytically, including the pre-exponential factor, using the theory of rare
events~\cite{CM10}. Therefore, long-time stochastic dynamics are adequately described by
a continuous-time Markov chain process on the graph introduced above. It can
be described by a Master Equation (ME), with the spectrum of the Master
Operator (MO) representing the soft modes of the spectrum of the Fokker-Planck
(FP) operator, the latter describing Langevin dynamics in $X$. Note that the
graph is naturally embedded into $X$, with its edges representing the most
probable transition trajectories (also known as optimal fluctuations or
instantons) that dominate the probabilities of the transition events. It is
intuitively natural that in the case when one follows the low-noise evolution
of higher-dimensional objects, the graph should be replaced by a
higher-dimensional object. Such an object is known as a {\em CW complex}, and
plays an important role in differential and algebraic topology; our
application deals with a special example of a CW complex, namely the
{\rm Morse complex} of a gradient flow.

In what follows we describe how a CW complex generalizes a notion of a graph
to higher dimensions. One can think about a graph as built in the following
way. We start with a set $K_0$ of the graph vertices, or $0$-dimensional cells;
they form the $0$-skeleton $K^{(0)}$. We further consider the set $K_1$ of
edges, or $1$-dimensional cells, with each edge being represented by a segment.
Edges are attached to the $0$-skeleton along their boundaries, represented by
pairs of points; this forms the $1$-skeleton $K^{(1)}$, which is the same as
a graph. Stated differently, if we stopped at this step, then we would be studying graphs. 
However, we can go
further and consider a set $K_2$ of $2$-cells (we omit ``dimensional'')
represented by $2$-disks $D^2$ and attach them to the $1$-skeleton along their
boundaries, represented by circles $S^1$, by means of maps $S^1 \to K^{(1)}$,
referred to as attaching maps. If we continue the procedure attaching $i$-discs
to the $(i-1)$-skeleton along their boundaries, represented by $(i-1)$-spheres
$S^{i-1}$ and eventually stop at the $m$-skeleton, as a result we build a space called an
$m$-dimensional CW complex. If a space $X$ is represented by a CW complex $K$,
we call the latter a CW-decomposition of $X$. We emphasize that within the described
terminology, a graph is nothing more than a $1$-dimensional CW complex. Also note that each
manifold of dimension $D$ can be triangulated, i.e., partitioned into a set of
triangles for $D = 2$, tetrahedra for $D = 3$, etc. A triangulation is the
simplest example of a CW-decomposition, with its structure being
transparent and obvious. For example, a $3$-dimensional manifold has its
$3$-skeleton as the manifold itself, the $2$- $1$- and $0$-skeletons are
represented by the unions of the tetrahedra faces, edges, and vertices,
respectively.

The concept of higher-dimensional current on a CW complex, which generalizes the
notion of current on a graph, is most naturally formulated in terms of
the chain complex of a CW complex. An $n$-chain is a linear superposition of $n$-cells 
with integer or real coefficients; the spaces of $n$-chains are denoted $C_n(K)$ or $C_n(K;
\mathbb{R})$ respectively. The boundary operator $\partial_n : C_n \to C_{n-1}$
maps an $n$-cell to a linear superposition of $(n-1)$-cells, completely determined by
the attaching maps. The integer coefficients 
characterize how many times (accounting for orientations) the boundary of the
$n$-cell covers the relevant $(n-1)$-cell. In our applications, we deal with
{\it regular} CW complexes, i.e., the ones for which the coefficients in the
aforementioned superpositions are $\pm 1$ or $0$. Note that a triangulation is a regular
CW complex. The inclusion $B_n \subset Z_n \subset C_n$, with $B_n = {\rm
im}(\partial_{n+1})$ and $Z_n = \ker(\partial_n)$ referred to as $n$-boundaries and $n$-cycles,
reiterates the intuitive and natural fact that the boundary of something is a
cycle, i.e., $\partial \circ \partial \, x = 0$ for any $x$ and in any dimension.
The aforementioned inclusion allows CW-homology to be defined by setting
$H_n(K) = Z_{n}(K)/B_{n}(K)$, or over real coefficients $H_n(K;\R) =
Z_{n}(K;\R)/B_{n}(K;\R)$. The importance of CW-homology is that it gives a
computationally accessible handle on the homology of the underlying space,
i.e., if $K$ is a CW-decomposition of $X$, then $H_i(X) \cong H_i(K)$, and
$H_i(X; \mathbb{R}) \cong H_i(K; \mathbb{R})$. With integer coefficients, this is valid for
the singular homology of $X$ also. With real coefficients, it works also for a broader class of
homology, e.g., bordism homology considered in \cite{CCK16b}.

The Morse CW complex (decomposition) is a very natural construction that
characterizes a gradient flow: the set $K_n$ of $n$-cells is represented by
critical points of Morse index $n$. The gradient flow generates a $n$-cell out
of a small neighborhood of the critical point, often referred to as the
unstable manifold, naturally embedded into $X$ and attached to the
$(n-1)$-skeleton; the latter is determined by the long-time behavior of the
flow. If we are looking at the low-noise, stochastic motion of
higher-dimensional cycles in the long-time limit, it can be reduced to a
discrete stochastic model on the Morse CW complex, similar to how it is
done in the simpler case of motion of points that is reduced to a Markov chain
model on a graph (the latter being actually represented by the $1$-skeleton of
the Morse CW complex). The details of this reduction are presented in
section~\ref{sec:Markov-chain}, here we provide a simple intuitive picture for
the $n = 2$ case, i.e., $1$-dimensional cycles create $2$-dimensional
trajectories.

The short-time relaxation of an initial $1$-cycle $\gamma: S^1 \to X$ forces it
to approach the $1$-skeleton.\footnote{In chemical physics this effect forms a
basis for a number of algorithms for finding transition states in molecules.}
This defines a $1$-chain $\bar{\gamma} \in C_1(K)$ of the Morse complex by
counting how many times (with the proper $\pm 1$ sign weights,
accounting for orientations) the relaxed cycle goes along the $1$-cells (edges)
of the $1$-skeleton, represented by a graph. Long-time dynamics are represented
by rare transitions when a fragment of the $1$-cycle associated with some given
edge goes over one of $2$-cells it is attached to, replacing the initial
fragment with the set of fragments that go over the rest of the edges the
$2$-cell is attached to. Such an effect is referred to as an elementary
transition, because it generalizes the notion of a jump over an edge in a standard continuous-time
Markov chain process. A sequence of elementary
transitions naturally creates a $2$-chain $\bar{\eta}$ that characterizes a
discrete counterpart of a stochastic trajectory. Averaging over stochastic
trajectories and dividing by the total time of the process we obtain a map
$Z_1(K) \to C_2(K; \mathbb{R})$. If in the $t \to \infty$ limit, we have
$t^{-1}\partial_2\bar{\eta} \to 0$, we refer to the situation as ``the
stochastic trajectories are statistically closed''. This condition implies $t^{-1} \bar{\eta}$
is a cycle, and thus gives rise to a homology class $[t^{-1}\bar{\eta}]$. 
In this way, we obtain a linear map $H_1(K) \to H_2(K; \mathbb{R})$, or in the general case a linear map
\begin{equation}
\label{flux-map-CW} \omega: H_{n-1}(K) \to H_n(K; \mathbb{R}),
\end{equation}
referred to as the flux map. Replacing the CW-homology of the Morse complex
with the homology of the underlying space $X$, we obtain a flux map of the form
considered in \cite{CCK16b} for a continuous setting of stationary or
periodic driving. We use the same notation for the both, since the flux map of
Eq.~\eqref{flux-map-CW} should reproduce its more general counterpart 
in the cases when the discrete stochastic
dynamics on a CW complex adequately approximates the continuous stochastic
dynamics on the underlying space.

Our formula for the average current density is given explicitly in terms
of subcomplexes of $K$. A {\em subcomplex} $L$ of $K$ is a CW complex in
its own right, together with an inclusion $L \subset K$ that preserves 
the relevant structure, i.e., attaching maps. A {\em spanning tree} (of dimension $d$)
is a subcomplex $T$ of $K$ so that $H_d(T) = 0$ and $H_i(T;\R) 
= H_i(K;\R)$ for all $i<d$. This should be thought of as a crude approximation
of $K$, since we have removed all $d$-dimensional `loops', but nothing
more. Every spanning tree acquires a weight as well, which
arises in the formula for current density and quantization. The
{\em weight} of a spanning tree $T$ is
\begin{equation}
  w_T = \theta_T^2 \prod_{\alpha \in T_d} e^{-\kappa^{-1} W_{\alpha}} \, ,
  \label{tree-weight}
\end{equation}
where the product is taken over all $d$-cells in $T$ and $\theta_T$
is the order of the torsion subgroup of $H_{d-1}(T)$. The number
$\theta_T$ is a measure of the difference between $H_{d-1}(T)$ and
$H_{d-1}(T;\R)$ and, so far as we can determine, has no physical 
meaning. In the analogy with resistive networks, each factor in 
this formula can be interpreted as the resistance of an edge, and therefore their 
product is the resistance of the network determined by $T$.

A {\em spanning co-tree} (of dimension $(d-1)$) is a subcomplex $L$ of $K$
for which $H_i(L;\R) = H_i(K;\R)$ for all $i<d$. These should be thought
of as even cruder approximations of $K$ than spanning trees, since they
can be given by brute truncation. In the case
when $K$ is one-dimensional, and hence just a graph, spanning trees coincide
with their classical counterparts, and spanning co-trees are just single vertices.
Spanning co-trees also have a weight given by
\begin{equation}
  \tau_L = c_L^2 \prod_{b \in L_{d-1}} e^{-\kappa^{-1} E_b} \, ,
  \label{weight-co-tree}
\end{equation}
where the product is taken over all $(d-1)$-cells of the co-tree, 
and $c_L$ is again given in terms of torsion subgroups of $H_{d-1}(X)$
and $H_{d-1}(L)$. The precise definition of $c_L$ is given in~\cite{CCK15b};
this definition sheds little insight physically and we omit it for
clarity.
The weights for both trees and co-trees fall out of constructing
pseudo-inverses, and so one should think of them as the necessary 
factors for the current formula to be valid.

\subsection{The Main Results: Higher-Dimensional Kirchhoff Theorems and Quantization of Stochastic Currents}
\label{sec:main results}

The Kirchhoff network problem for graphs asks, given voltages
in a resistive network, find the unique current flowing through the
network. The first mathematically
rigorous treatment of this problem was given by Hermann Weyl in
1923~\cite{W23}, and an algebraic topological proof appeared
shortly thereafter~\cite{NS61}. We aim to generalize this problem
and its solution to arbitrary CW complexes. 

We are passing from one description of dynamics (a smooth, cohomological one in
terms of differential forms) to its dual (a discrete, homological picture in
terms of chain complexes).  The main object of study will now be the chain
complex $C_*(K)$, instead of the Grassmann algebra, and so the operator ${\cal
Q}$ is replaced with the differential $\partial$ on $C_*(K)$. The reduced
dynamics on $K$ requires us to only know the values of the potential $V$ at the
finite set of critical points; set $E_{i} := V(i)$ and $W_{\alpha} :=
V(\alpha)$ for each critical point of index $(n-1)$ and $n$, respectively.
Finally, a {\em driving protocol} keeps track of this data; it assigns to each
index $(n-1)$-critical point $i$, the real number $E_i$, and to each index
$n$-critical point $\alpha$, $W_{\alpha}$. We are concerned with the non-stationary
case, so the driving protocol will be a function of time.

The first quantities to translate to the discrete setting are the
SFP operator and the current density.
By constructing an adjoint $\partial^{\dagger}$ to the boundary operator
$\partial$ in the obvious fashion, we obtain the discrete counterpart
to the continuous SFP equation (Eq.~(25) of in \cite{CCK16b})
\begin{equation}
  \frac{d \rho}{d \tau} = - \partial {J} \, , \quad 
  {J} =  e^{-\kappa^{-1} \hat W(t)} \partial^{\dagger}
  e^{\kappa^{-1} \hat E(t)} \rho \, .
  \label{SFP-eq-discrete}
\end{equation}
We use $\rho$ for the distribution in the discrete, CW setting 
to distinguish it from the continuous distribution $\varrho$.

The main result of this paper concerns an equivalent formula for the
current density $J$ which is computable, unlike the 
purely formal definition given above. The form combines the solution
to Kirchhoff's higher network problem and the Boltzmann distribution 
in higher dimensions. Kirchhoff's problem can be solved using
pseudo-inverses, even in the discrete case, with an operator
analogous to $A_{V(\tau),\kappa}$ of \cite{CCK16b}. 
Here, the operator $A_{\hat W(\tau),\kappa}$
acts on $(d-1)$-boundaries to produce $d$-chains. It is
uniquely determined by the following two conditions:
\begin{equation}
  \label{define-A-prop}
  \partial A_{\hat W(t), \kappa} y = y,  \quad
  \quad (A_{\hat W(t),\kappa} y , z )_{\hat W(t), \kappa} = 0\, ,
\end{equation}
for any $y \in B_{d-1}(K;\R)$ and $z \in Z_d(K;\R)$. The modified 
inner product is defined to be
\begin{equation}
   \label{eq:mod-in-prod}
    (a , b)_{\hat W(t), \kappa} = (e^{\kappa^{-1}\hat W(t)} a , b) \,,
\end{equation}
for any $a$ and $b$ in $C_{d-1}(K;\R)$.
These two equations are the discrete analogues of Eqs.~(34) and (33)
from \cite{CCK16b}, respectively.

The Boltzmann distribution $\rho_B([\gamma]; \hat E(\tau), \kappa)$ 
is also determined by a pair of related
constraints. From a topological viewpoint, the
Boltzmann distribution is equivalent to finding a preferred cycle
representative for any homology class, as the latter is a quotient of the
former.  This linear operator is subject to:
\begin{enumerate}
  \item the cycle $\rho_B([\gamma])$ lies in the same homology class 
    as $[\gamma]$ itself; that is, $[ \rho_B([\gamma]) ] = [\gamma]$, and
  \item The inner product $( \rho_B([\gamma];\hat E(\tau), \kappa), \partial \alpha)_{\hat E(\tau), \kappa} = 0$ 
    for any $\alpha \in C_{d-1}(X;\R)$.
\end{enumerate}

These operators are defined in subsections~\ref{sec:Kirchhoff-higher-tree}
and~\ref{sec:Kirchhoff-higher-cotree}. The two sets of conditions defining
$A$ and $\rho_B$ are dual. This symmetry allows a single, universal
construction to be given for both. In this case, the construction is known as
a {\it pseudo-inverse} \cite{MR1987382}. This is a one-sided inverse (see the first condition in both),
subject to an additional constraint (the second condition in both). This
allows us to write both operators as a sum indexed over subcomplexes of $K$,
giving these physical constructions intrinsic geometric meaning. 

There is a tremendous amount of topological and combinatorial
information about the Morse complex, and hence the stochastic
dynamics of the process, contained in these two operators. They
are both written in terms of higher dimensional version of
spanning trees and vertices generalized from graphs. 
Together, they combine to yield
\begin{equation}
   J([\gamma], \kappa, \tau) = A(\tau) \dot \rho_B([\gamma],\tau) \, ,
  \label{J-A-rhoB}
\end{equation}
a computable version of the current density. This allows us to write down the
explicit integral representation for the average current density
as 
\begin{equation}
  \label{Q-A-rhoB}
  Q([\gamma], \kappa, u(\tau)) = \int_0^1 d \tau J(\tau) \, .
\end{equation}

In order to obtain meaningful results and have some hope of computing
Eq.~\eqref{Q-A-rhoB}, we must restrict the parameters we allow for 
the process. Specifically, we narrow our parameters to those which 
satisfy a generic, non-degeneracy condition. A driving protocol is
{\em very non-degenerate} if, at any moment in time, either
$\hat E$ or $\hat W$ satisfies one of the following (in place of $f$)
\begin{equation}
  \sum_{s \in S} f(s) \neq \sum_{t \in T} f(t) \, , \quad \mbox{ for all } S \, , T
  \label{very-non-degen}
\end{equation}
where $S$ and $T$ are subsets of the $(d-1)$-cells (for $\hat E$) or $d$-cells
(for $\hat W$). This condition is a higher analogue for a function to
be injective, or one-to-one, and recovers this notion if $S$ and $T$ are 
taken to be singletons. This condition, and therefore this subspace of
driving protocols, is generic and so there is little lost by this restriction.
We shall restrict to very non-degenerate driving protocols for the remainder
of this text.

It is easy to see that a non-stationary driving protocol can be arranged
to alternate between segments where either $\hat E$ or $\hat W$ (or both)
satisfy Eq.~\eqref{very-non-degen}; segments for which $\hat E$ satisfies
this are known as {\em type E}, and those for $\hat W$ are {\em type W}.
The difference between these two segments is visible in the stochastic
dynamics. On segments of type of $E$, the evolving cycle will tend
to a configuration of minimal `energy', i.e., it will minimize 
$\sum_b E_b$ over the cycle, and be entirely supported on a uniquely
determined spanning co-tree. On a type $W$ segment, the cycle can move
quite substantially across the complex, but in a way
so that its motion is restricted to a specific spanning tree. In fact, it will move
along the spanning tree from one spanning co-tree to another, with all
of these structures being uniquely determined in the low-temperature, 
adiabatic limit. This motion of moving from co-tree to co-tree along trees
is in direct analogy with (and should be thought of as) a particle moving 
on a graph by jumping from vertex to vertex along a (traditional) spanning tree.
On the type $E$ segments, where the cycle is restricted to a single spanning
co-tree, there is obviously no current generated. On the type $W$ segments,
where the cycle is transitioning from one spanning co-tree to another,
current is generated and the values this current takes are determined
by the relevant structures involved.
Analyzing this motion, we are able to write down an explicit formula
for the average current in these limits, and conclude that the
coefficients obtained are rational. That is, the main result of this
text is the following quantization result:
\begin{equation}
  \lim_{\kappa \ra 0}  \lim_{\tau_D \ra \infty} Q([\gamma],\kappa, \tau) 
  \in H_{d}(K;\Q) \, .
  \label{rational-quant-Q}
\end{equation}

\section{Markov Chain Models for Empirical Currents in the Weak-Noise Limit}
\label{sec:Markov-chain}

Similar to the case of standard empirical currents, considered in some detail
in \cite{CCMT09}, the problem of higher-dimensional empirical current
generation for moderate topological driving that satisfies the Morse-Smale
condition, in the weak-noise limit can be reduced to Markov Chain (MC) models.
This reduction will be addressed with some detail elsewhere. Instead, in this
section we introduce the MC models using intuitive physical arguments, with the
relevant transition (hopping) rates considered as phenomenological constants, in the case
when the driving is topological, which means that the deterministic
(advection) component $u$ in the Langevin equation is given by $u^{j}(x)=g^{jk}(x)F_{j}(x)$
with $\partial_{i}F_{j}(x)-\partial_{j}F_{i}(x)=0$. We can recast this in an
equivalent form as ${\cal Q}(F_{j}\Theta^{j})=0$. Moderate driving means that the
critical subspace, which contains of points with $F(x)=0$, is represented by a
finite set of isolated non-degenerate (the matrix $\partial_{i}F_{j}(x)$ is not
degenerate) critical points. Recall that the number of unstable modes near a critical
point, i.e., the number of negative eigenvalues of $\partial_{i}F_{j}(x)$, is
referred to as the Morse index of a critical point.

We will also assume the Morse-Smale (MS) condition \cite{Hirsch97} to be
satisfied. Note that the MS condition is usually formulated for gradient flows,
however, all results immediately apply to our case of topological driving
(i.e., locally gradient flow) with the simple critical structure of moderate
driving, as described above. This follows from the fact that within the cells,
associated with critical points, the flow is gradient, since the cells are
contractible by definition. Let $K_{0},K_{1},\ldots, K_{m}$ be the (finite) sets
of critical points with a given Morse index. The most important property of a
flow satisfying the MS condition is that it provides a cellular (Morse)
decomposition of the configuration space $X$,
%\footnote{It should be stressed
% that there is certain controversy in the differential topology literature,
%  regarding the described property and some related properties within the  picture
%  presented below. Most recently it has been demonstrated \cite{KQ09} that the
%above picture holds when the metric $g$ is locally flat in the neighborhoods of
% the critical points. It is unclear how important the local flatness requirement
%(that does not appear physical) is. Anyway, if the requirement turns out to be
%important, our consideration can provide certain implication on physical
% observables.} 
in a way that each critical point $y\in K_{n}$ provides a
$n$-dimensional cell $W_{y}$, produced by the unstable manifold of $y$ (i.e.,
$W_{y}$ consists of all points covered by the flow that starts at $y$ in all
unstable directions). We will fix the orientations for all cells in some
arbitrary way by choosing the orientations of the unstable subspaces at the
corresponding critical points \cite{Qin1},\cite{KQ09},\cite{Qin2}. The cells can be properly compactified (by
inspecting the infinite-time flow behavior) so that the border $\partial W_{y}$
is represented by a union of a number of cells related to critical points of
index $(n-1)$. If $W_{y'}\subset \partial W_{y}$ for some critical point $y'\in
K_{n-1}$, the intersection $p_{yy'}$ of the unstable manifold of $y$ with the
stable manifold of $y'$ is represented by a flow trajectory that starts at $y$
and terminates at $y'$ in infinite time. The set of such trajectories is
denoted $P_{yy'}=\{p_{yy'}\}$. With each trajectory $p_{yy'}\in P_{yy'}$ we can
associate a binary variable $s(p_{yy'})\in \mathbb{Z}_{2}$, so that
$s(p_{yy'})=0$ or $s(p_{yy'})=1$, when the orientation of $W_{y'}\subset
\partial W_{y}$ matches or mismatches the orientation of $W_{y}$, respectively.
The set $C_{\bullet}(K) = \mathbb{Z}(K_{\bullet})$ of abelian groups, referred to as the groups
of chains is equipped with a set $\partial_{\bullet}:C_{\bullet}(K) \ra
C_{\bullet -1}(K)$ of boundary operators, defined by
\begin{equation}
\label{define-boundary-oper} \partial_{\bullet}y= \sum_{y'\in K_{\bullet-1}}\sum_{p\in P_{yy'}}(-1)^{s(p)}y', \;\;\; {\rm for} \;\; y\in K_{\bullet},
\end{equation}
so that $(C_{\bullet}(K),\partial_{\bullet})$ forms a complex,
referred to as the Morse complex, associated with the described above Morse
decomposition.

The above the Morse decomposition provides a set of
bipartite graphs $G_{n}=(K_{n+1},K_{n};E_{n})$ for $n=0,1,\ldots,m-1$ with the
edge sets $E_{n}=\bigcup_{(y,y')\in K_{n+1}\times K_{n}}P_{yy'}$. Stated in
words, the two types of nodes in bipartite $G_{n}$ are given by critical points
of index $(n+1)$ and $n$, whereas the edges represent the flow trajectories
that start at $K_{n+1}$-nodes and terminate at $K_{n}$-nodes. The graph $G_{n}$
plays a crucial role in the description of slow stochastic dynamics of
$n$-cycles and corresponding $n$-dimensional current generation in the
weak-noise limit of moderate topological driving that satisfies the MS
condition. In the above limit the problem can be reduced to a standard MC
process on an infinite oriented graph ${\cal G}_{n}=({\cal N}_{n},{\cal
E}_{n})$. We will first build the graph ${\cal G}_{n}$ and define a MC process
on it, and after that provide a physical motivation for such a reduction. The
reduction (including deriving explicit expressions for the hopping rates) will
be addressed elsewhere.

We will denote the nodes of the two types $\alpha\in K_{n+1}$ and $j\in K_{n}$
by Greek and Latin letters, respectively. The nodes of ${\cal G}_{n}$ that
describe the metastable states of slow stochastic dynamics are given by ${\cal
N}_{n}={\rm ker}\, \partial_{n}\subset C_n(K)$, which means that a
metastable state ${\bm n}\in {\cal N}_{n}$ is an integer vector with the
components $n_{j}\in \mathbb{Z}$ and $j\in K_{n}$. The (directed)
edges (that belong to ${\cal E}_{n}$) correspond to elementary hopping events.
The edges are partitioned in groups labeled by $(j,\alpha;p_{\alpha j})$ with
$j\in K_{n}$, $\alpha\in K_{n+1}$, and $p_{\alpha j}\in P_{\alpha j}$, i.e.,
${\cal E}_{n}=\bigsqcup{\cal E}_{j\alpha,p_{\alpha j}}^{n}$. All elementary
hopping processes that go along the edges of the same group have the same
hopping rate denoted by $k_{\alpha j,p_{\alpha j}}$. This means that an
elementary process has a beginning $j\in K_{n}$ related to a metastable state,
a ``transition' state $\alpha\in K_{n+1}$ and a ``pathway'' $p_{\alpha j}\in
P_{\alpha j}$ that connects the beginning to the ``transition'' (a more
adequate physical explanation will be given later in this section). To complete
the description of our MC model we should describe all edges that outgo from
any given node. Consider an arbitrary node ${\bm n}\in {\cal E}_{n}$ and a
group ${\cal E}_{j\alpha,p_{\alpha j}}^{n}$ of edges labeled by
$(j,\alpha;p_{\alpha j})$. If $n_{j}=0$ there are no edges in the group that
outgoes from ${\bm n}$. If $n_{j}\ne 0$ there is exactly one edge that outgo from
${\bm n}$. The node ${\bm n}'$, where this unique edge is incoming, can be described
as follows:
\begin{equation}
\label{elementary-process} 
\begin{aligned}
  n'_{k}&=n_{k}-\sum_{r\in P_{\alpha k}}(-1)^{s(p_{\alpha j})+s(r)}n_{j}, \\
  {\bm n}'&={\bm n}-(-1)^{s(p_{\alpha j})}\partial_{n+1}{\bm n}.
\end{aligned}
\end{equation}
The sign factor $(-1)^{s(p)+s(r)}=\pm 1$ reflects the relative orientation of
the cells $W_{j}\subset W_{\alpha}$ and $W_{k}\subset W_{\alpha}$ with respect
to $W_{\alpha}$. The second relation in Eq.~(\ref{elementary-process}) recasts
the first relation using the definition of the boundary operator
[Eq.~(\ref{define-boundary-oper})] and clearly demonstrates the fact that the
homology class of the state $[{\bm n}]$ is preserved in the described MC
process.

We are now in a position to present the physical motivation that stands behind
the described above MC process on the graph ${\cal G}_{n}$. The physical picture is
based on the time scale separation. The initial (fast) relaxation occurs on a
finite time scale in the $\kappa\to 0$ limit. During the initial relaxation, the
initial cycle evolves essentially deterministically and as a result gets
aligned with the $n$ dimensional cells $W_{j}$, while performing weak fluctuations in
a small neighborhood $U_{n}X\supset K^{(n)}=\bigcup_{j\in K_{n}}W_{y}$ of the
the $n$-skeleton $K^{(n)}$ associated with the Morse decomposition of $X$.
After the fast relaxation is completed, the evolution of
the cycle is represented by a sequence of rare events during which the relevant
fragments of the cycle move along the $(n+1)$-cells $W_{\alpha}$. Stated more
formally, most of the time the cycle ($n$-dimensional surface in $X$) is
restricted to $U_{n}X$, whereas while performing the rare transitions the cycle
is restricted to a broader subspace $U_{n+1}X\supset U_{n}X$. Since a
transition is rare, it occurs as an optimal fluctuation (that maximizes the
probability of the event) according to the following scenario. A fluctuation,
represented by a Langevin vector field $\xi$ occurs in the vicinity of a
critical point $j\in K_{n}$ that drags a small fragment of the cycle
$\zeta:N\to X$ located near $j\in X$ (the fragment is typically folded) along
the path $p_{\alpha j}\in P_{\alpha j}$ to a critical point $\alpha\in K_{n+1}$
that labels a cell $W_{\alpha}$ attached to the cell $W_{y}$. Once the fragment
passes the critical point no more noise is required to complete the transition:
the fragment, that has been dragged, acquires intersections with the paths
$p_{\alpha k}\ne p_{\alpha j}$ and, on the time scale of fast relaxation, the
intersections, typically represented by isolated points, will relax to the
critical points $k\in W_{n}$ going along the paths $p_{\alpha k}$, mentioned
above. As a result the whole bigger fragment of the cycle that used to be
located in a small neighborhood of $W_{j}$ will locate itself in the small
neighborhoods of the cells $W_{k}$, for all $k\in K_{n}$ such that $W_{k}\subset
\partial W_{\alpha}$. Note that during the whole transition the whole fragment
moves along a small neighborhood of the ``transition'' $(n+1)$-cell
$W_{\alpha}$.

The reduced variables ${\bm n}\in {\cal N}_{n}$ that characterize the
metastable states and form the nodes of the graph ${\cal G}_{n}$ introduced
above are of topological nature. They can be described as follows. Consider a
metastable configuration of our cycle $\zeta$, when it is located in the small
neighborhood $U_{n}X\supset K^{(n)}$ of the $n$-skeleton of $X$. We further take
a critical point $j\in K_{n}$, look at all components of our cycle located near
$W_{j}$ (typically the cycle is folded) and define $n_{j}$ as the number of
times the cycle appears in the neighborhood with the same orientation as
$W_{j}$ minus the number of times it appears with the opposite orientation.
This can be formulated more precisely by defining $n_{j}$ as the intersection
index of $\zeta:N\to X$ with the stable manifold of the critical point $j$. The
transverse size of $U_{n}X$, where a metastable cycle is concentrated, is
$\sim\sqrt{\kappa}$. Therefore, if the correlation length of the random field
$\xi$ is large compared to $\sqrt{\kappa}$ (which will always happen for low
enough noise), all components of the cycle will be dragged through the
transition state together. On the other hand, if the correlation length is
small compared to the cell sizes, the elementary transitions will be restricted
to individual cells $W_{\alpha}$. This rationalizes the description in terms of
the reduced topological variables ${\bm n}$, as well as the structure of the
elementary transitions, described by Eq.~(\ref{elementary-process}).

The fact that the obtained MC process occurs on an infinite graph reflects the
field-theory nature of the original problem. We have demonstrated, however,
that the moments of the current distributions, including the stationary current
can be calculated using spectral decompositions for operators acting in the
spaces of states (referred to as the spaces of correlation functions)
associated with finite-dimensional manifolds. Naturally such a reduction is
inherited by the MC models, i.e., the moments of the current distributions can
be calculated by using spectral decompositions in finite-dimensional vector
spaces that describe the corresponding correlation functions.

Let $\mu({\bm n})$ be the probability distribution that describes the MC
process on ${\cal G}_{n}$, where it satisfies the appropriate master equation.
We start with the stationary current that can be expressed in terms of a
correlation function (reduced distribution) that belongs to $\mathbb{R}(K_{n})$
and is described by a vector with the components
\begin{equation}
\label{define-CF-discrete} \rho_{i}=\sum_{{\bm n}}n_{i}\mu({\bm n})
\end{equation}
By calculating the time derivative $\dot{\rho}_{j}$ via applying the full
master equation for $\mu({\bm n})$ we obtain, after some straightforward
transformations, a closed equation for $\rho$ that has a natural form
\begin{equation}
\label{FP-CF-discrete} 
\begin{aligned}
  \dot{\rho}_{i} &=\sum_{\alpha\in K_{n+1}}\sum_{j\in K_{n}}^{j\ne i}\sum_{p\in P_{\alpha j}}\sum_{r\in P_{\alpha i}}k_{\alpha j,p}(-1)^{s(p)+s(r)}\rho_{j} \\
  &-\sum_{\alpha\in K_{n+1}}\sum_{p\in P_{\alpha i}}k_{\alpha i,p}\rho_{i} \, .
\end{aligned}
\end{equation}

The only trick used in deriving Eq.~(\ref{FP-CF-discrete}) is introducing the axillary correlation functions
\begin{equation}
\rho_{i}^{(j)}=\sum_{{\bm n}}^{n_{j}\ne 0}n_{i}\mu({\bm n}),
\end{equation}
and appreciating $\rho_{i}^{(i)}=\rho_{i}$.

\section{Generated Higher-Dimensional Currents in a Periodically Driven System on a CW complex}
\label{sec:generated-currents-CW}

In this section, we develop the necessary tools and give an explicit integral
representation for the average current density generated per period of 
driving protocol. Our method relies on the reduction to the Morse CW complex $K$ of 
$X$ discussed in the previous section. This is a generalization of the 
one-dimensional case, given by writing the SFP operator in a geometric
fashion on the chain complex of $K$. 
We then write the integral representation of the generated
current in terms of two key components, the solution to Kirchhoff's network problem
and a higher-dimensional version of the Boltzmann distribution,
both of which are familiar from the classical, one-dimensional case. In
subsection~\ref{sec:Kirchhoff-higher-tree}, we discuss Kirchhoff's network
problem and its solution, given by application of the aforementioned theory of
pseudo-inverses. The second component is the Boltzmann distribution, studied in
subsection~\ref{sec:Kirchhoff-higher-cotree}. After obtaining both expressions
in higher dimensions, we combine them to yield the desired result
in subsection~\ref{sec:low-T-adiabatic-explicit}.

%Our formulation of Kirchhoff's network problem and our methods of proof
%generalize an algebraic topological proof given in [NS].
%
%
%Using the theory
%of pseudo-inverses, we are able to solve both at the same time. Our formulation of the
%network problem follows the algebraic topological proof given in~[NS]. After
%obtaining both expressions in higher dimensions, we combine them to yield
%the desired result in subsection~\ref{sec:low-T-adiabatic-explicit}.

\subsection{Integral Representation for the Average Generated Current}
\label{sec:integral-formula}

The first goal is to express Eq.~\eqref{FP-CF-discrete} in terms of a
supersymmetric Fokker-Planck operator on the CW complex $K$. 
%This version of the
%SFP serves as the discrete analog of Eq.~\eqref{SFP-eq-continuity}.
We begin by constructing an adjoint to the boundary map $\partial: C_n(K) \ra C_{n-1}(K)$ 
defined in Eq.~\eqref{define-boundary-oper}. The boundary map should be
thought of as a discrete analogue of the divergence operator from calculus,
and so the adjoint map $\partial^{\dagger}$ should be thought of as a 
gradient operator. It is given by
\begin{equation}
  \partial^{\dagger} y' = \sum_{y \in S_n} \sum_{p \in P_{y y'}} (-1)^{s(p)}
  y \quad y \in S_{n-1} \, .
  \label{define-adjoint-oper}
\end{equation}
Both $\partial$ and $\partial^{\dagger}$ are useful operators for our 
formulation since their definitions are orientation-independent. Furthermore,
they provide concise definitions of the average current density.

All that remains for the SFP operator are the
rates of the process. We are focused with the case when the
transition rates satisfy detailed balance at each moment in time. We 
write them in Arrhenius form
\begin{equation}
  \label{rates-Arr-form}
  k_{i \alpha, p}(t) = e^{-\kappa^{-1}(W_{\alpha}(t) - E_i(t))} \, ,
\end{equation}
where $k_{i \alpha, p}$ is the rate of hopping along the pathway $p$,
starting at cell $i$, and transitioning over $\alpha$. We encorporate
these rates as linear operators by defining diagonal matrices 
$\hat E(t) = \mathrm{diag}(E_1(t), \ldots, E_{|K_{d-1}|}(t))$
and $\hat W(t) = \mathrm{diag}(W_1(t),\ldots, W_{|K_d|}(t))$. 
These definitions
combine to allow us to define the SFP operator on $C_{\bullet}(K)$ as
\begin{equation}
  \hat H(\gamma(t)) := \partial e^{-\kappa^{-1} \hat W(t)} \partial^{\dagger} e^{\kappa^{-1} \hat E(t)} \, .
  \label{define-FP-oper}
\end{equation}
It is a straightforward application of the above definitions to verify 
that $\hat H \rho$ coincides with the r.h.s. of Eq.~\eqref{FP-CF-discrete}. Therefore,
we succinctly recast Eq.~\eqref{FP-CF-discrete} as 
\begin{equation}
  \label{SFP-eq-succinct}
  \hat H(t) \rho(t) = \dot \rho(t) \, ,
\end{equation}
resembling the Fokker-Planck equation from Langevin dynamics. Furthermore,
this serves as a discrete analog to the SFP equation considered in 
\cite{CCK16b}.

%Given an initial distribution $\rho(0)$ and driving protocol $\gamma$, the 
%stochastic current density averaged over the driving protocol is given
%by 
%\begin{equation}
%  \mathbf J(\gamma(t)) = e^{-\kappa^{-1} \hat W(t)} \partial^{\dagger} e^{\kappa^{-1} \hat E(t)} \, ,
%  \label{define-J}
%\end{equation}
%where $\rho(t)$ is the unique solution to the Fokker-Planck equation, given by
%\begin{equation}
%  \rho(t) = \hat T \mathrm{exp} \left( \int_0^t dt' \hat H(\gamma(t')) \right) \rho(0)\, ,
%  \label{formal-soln}
%\end{equation}
%where we use the time-ordered exponential $\hat T \mathrm{exp}$. This definition of
%$\mathbf J$ is the discrete analog of the continuous version given in
%Eq.~\eqref{SFP-eq-continuity}.

We are primarily interested in the average current density over many 
periods of the driving protocol. The average current density $Q$
is given by
\begin{equation}
  Q(\gamma(t)) = \frac1N \int_0^{N \tau_D} dt \, J(t) \, , \quad N \gg 0 .
  \label{define-Q}
\end{equation}

There is an obvious adiabatic theorem which holds true in this generalized setting. It
has been treated mathematically in~\cite{CCK16}, and states that a 
periodic solution to the SFP equation exists and is unique. Furthermore, in
the adiabatic limit, this solution tends to the Boltzmann distribution. 
The remainder of this section is devoted to showing Eq.~\eqref{J-A-rhoB}
is valid.  An explicit
formula for both $A$ and $\rho^B$ will be given in the
next two subsections.
Combining these two pieces, we obtain the desired formula
\begin{equation}
  Q(\gamma(t)) = \int_0^1 d \tau A(\tau) \dot \rho^B(\tau) \, , \quad \tau_D \ra \infty \, ,
  \label{Q-final}
\end{equation}
where we have expressed the quantities in terms of dimensionless time $\tau = t/\tau_D$.

\subsection{Higher-Dimensional Network Kirchhoff Theorem}
\label{sec:Kirchhoff-higher-tree}

The classical Kirchhoff problem aims to find the unique current
in a resistive network given some initial current. The higher dimensional
network problem asks the analogous question for CW complexes of 
arbitrary dimension. 

The higher dimensional Kirchhoff problem, i.e. constructing a solution
$A = A_{\hat W(t), \kappa}$ to Eq.~\eqref{define-A-prop}, is solved by introducing
the discrete counterpart to the continuous pseudo-inverse
introduced in \cite{CCK16b}, which acts on $(d-1)$-boundaries to
produce $d$-chains. 
We will write down the operator $A$ explicitly in terms of higher dimensional
spanning trees, introduced in subsection~\ref{sec:CW-Morse-lowtemp} (we
omit the dependence of $A$ on $\hat W(t)$ and $\kappa$ in what follows). 
Fix a spanning tree $T$, and define an operator $A_T$ as follows. By
definition, every boundary in $X$ is the boundary of something in $T$. Or in
symbols, for every $b = \partial\alpha \in C_{d-1}(X;\R)$, there is a unique
$d$-chain in $T$, $A_T^b$, so that $\partial A_T^b = b$. Define $A_T(b) =
A_T^b$.  The solution to Kirchhoff's higher dimensional network theorem is then
given by taking a weighted sum of such operators, indexed over the set of all
spanning trees. The final form the operator takes is
\begin{equation}
  A = \tfrac{1}{\Delta} \sum_T w_T A_T \, ,
  \label{define-A}
\end{equation}
where $w_T$ is the weight of a tree, as defined in subsection~\ref{sec:main results}, and
$\Delta = \sum_T w_T$. It is straightforward to verify that this definition
of $A$ verifies the criteria of Eq.~\eqref{define-A-prop}. The first 
property follows directly from the definition of $A_T$, and the second follows
by noting that $A$ is self-adjoint with respect to the inner product of
Eq.~\eqref{eq:mod-in-prod}.

The operator $A_T$ has been given in a geometric fashion in terms of unique
chains in the spanning tree $T$. This is a direct generalization of the Kirchhoff
theorem for graphs, or one-dimensional complexes. In this case, the operator
$A_T$ sends a boundary, say given by a difference of vertices $j-i$,
to the unique path in the (ordinary) spanning tree $T$ from $i$ to $j$. In
fact, the unique current in a resistive network, i.e. a solution to Kirchhoff's
problem on graphs, is given by Eq.~\eqref{define-A}.

\subsection{Higher-Dimensional Boltzmann Distributions and Kirchhoff Co-Tree Theorem}
\label{sec:Kirchhoff-higher-cotree}

The other key ingredient needed for Eq.~\eqref{Q-final} is the higher dimensional
Boltzmann distribution.  This is a direct generalization of the classical
distribution from statistical mechanics. 

The higher dimensional Boltzmann problem (also known in topology as the Hodge decomposition
problem) is to find an explicit cycle representative for any homology class.
The Boltzmann distribution can be written explicitly in terms of spanning co-trees, 
introduced in subsection~\ref{sec:main results}. By their definition,
within each co-tree we can find a unique cycle representative for each homology class
of $X$. Therefore, for each spanning co-tree $L$, define $\psi_L([\gamma])$ to be the unique
cycle representative of $[\gamma]$ in $L$. We then take a weighted sum over these operators
to obtain
\begin{equation}
  \rho_B = \tfrac{1}{\Lambda} \sum_L \tau_L \psi_L \, ,
  \label{define-rhoB}
\end{equation}
where $\tau_L$ is the weight of the spanning co-tree $L$ as in
subsection~\ref{sec:main results} and $\Lambda = \sum_L \tau_L$.
As previously mentioned, this problem (and its solution) are dual
to the Kirchhoff problem, so the same proof used there applies here.
A rigorous proof of this fact without duality 
was given in~\cite{CCK15b} using pseudo-inverses. 

An important case to keep in mind is the familiar one of a graph, or one-dimensional CW complex.
In this case, the spanning co-trees are given by the vertices, 
$c_L = 1$, and each vertex $x$ is then
weighted with $\tau_x = e^{-\beta E_x}$. In this case, Eq.~\eqref{define-rhoB} becomes
\begin{equation}
  \rho_B = \frac{\sum_{x \in X_0} e^{-\beta E_x} \cdot x}{\sum_{x \in X_0} e^{-\beta E_x}} \, ,
  \label{classical-rhoB}
\end{equation}
reproducing the famous distribution.

\subsection{Low-Temperature Adiabatic Limit: Explicit Expression for the Generated Current}
\label{sec:low-T-adiabatic-explicit}
 We are now in a position to verify Eq.~\eqref{Q-final}. The equivalence
 between the expressions given in Eq.~\eqref{SFP-eq-discrete} and Eq.~\eqref{J-A-rhoB}
 can be seen by a straightforward argument. Consider the set of all $d$-chains
 $\nu \in C_d(X;\R)$ such that
 \begin{equation}
   \partial\nu = \dot\rho\, ,  \quad  ( \nu, z)_{\hat W(\tau), \kappa} = 0 \,,
   \label{J-equivalence}
 \end{equation}
 for every $z \in Z_{d}(K)$.
 First, it is clear that any $\nu$ satisfying these two conditions must be unique.
 If $\nu$ and $\nu'$ were both solutions, then the first condition of Eq.~\eqref{J-equivalence}
 implies $\partial(\nu - \nu') = 0$, so that $\nu - \nu'$ would be a cycle. The
 second condition then implies $\nu - \nu'$ to be orthogonal to itself, forcing
 $\nu = \nu'$. 
 It remains to show that the each of the aforementioned expressions
 satisfy these two criteria, after which the proof shall be complete. For 
 Eq.~\eqref{SFP-eq-discrete}, the two conditions are verified in precisely the same
 manner as in the continuous case (Eqs.~(56) and (57) in \cite{CCK16b}). 
 For Eq.~\eqref{J-A-rhoB}, both conditions are readily verified by the definition
 of $A$ in Eq.~\eqref{define-A-prop}.

\section{Low-Temperature Adiabatic Limit: Quantization of Generated Currents}
\label{sec:quantization-CW}

We now state and prove the main result on the current generated under a
multi-dimensional Langevin process: rational quantization. The quantization
relies on restricting to the  generic set of very non-degenerate driving
protocols, defined in subsection~\ref{sec:main results}.  This restriction
allows for a clear analysis in terms of two distinct types of segments of the
driving protocol. We take the adiabatic limit for the remainder of this
section, so that $N \ra \infty$.

On a segment of type $W$, only the $W$ parameters are changing, and the functional
$E$ is constant. With these parameters, the evolving fragment will remain fixed
and no current will be generated. This can be seen explicitly in Eq.~\eqref{Q-final}, 
by noting
that if $E$ is fixed, then $\dot \rho_B \ra 0$ and $\kappa \ra 0$:
\begin{equation}
  \lim_{\kappa \ra 0} Q = \int_a^b d\tau A \dot \rho_B = 0\, .
  \label{quant-type-W}
\end{equation}

On a type $E$ segment, the functional $E$ is varying and $W$ is fixed. 
It is with these parameters that average current density can be generated. Even
more, on an individual type $E$ segment, the average current density is 
determined by the driving protocol at the end points. 
In the limit as $\kappa \ra 0$, the linear map $A \ra A_T$, where
$T$ is determined uniquely to be the tree which minimizes
\begin{equation}
  \sum_{\alpha \in T} W(\alpha) \, .
  \label{unique-T}
\end{equation}
Therefore, in the low-noise limit, we have
\begin{equation}
  \begin{aligned}
    \lim_{\kappa \ra 0} Q &= \lim_{\kappa\ra 0} \int_c^d d\tau A \dot \rho_B \\
    &= A_T \int_c^d d \tau \dot \rho_B \\
    &= A_T (\psi_{L(d)} - \psi_{L(c)}) \, ,
  \label{quant-type-E}
\end{aligned}
\end{equation}
where $L(c)$ and $L(d)$ are the unique spanning co-tree at their corresponding
values of driving protocol. As previously mentioned, each of these
individual pieces are weighted with only rational numbers. Summing over
all alternations of segments, the average current density will only take 
on rational numbers or the value 0. Therefore, the average current density
will be a rational number in the low-temperature, adiabatic limit, completing
the proof.

Although the process we have been describing throughout this text is
inherently stochastic, the low-temperature, adiabatic limit allows us
to provide an intuitive, albeit overly simplified, picture from the viewpoint
of average current density. Suppose the parameters are varying by
starting at a segment of type $E$, then type $W$, then back to $E$. On 
each type $E$ segment, there is a preferred spanning co-tree; call
them $L$ and $L'$, respectively. On the type $W$ segment, there is 
a unique spanning tree $T$. By the defining properties of spanning
trees, they contain all spanning co-trees; in particular, $T$ contains
both $L$ and $L'$. On the initial type $E$ segment, the evolving cycle
tends to $L$, where it will remain with an overwhelming probability. 
As the driving protocol transitions to the type $W$ segment, the cycle
will begin to fluctuate in the spanning tree $T$. It will continue evolving
in $T$, with its motion restricted to the subcomplex $A_T^{L - L'}$,
until it gets to $L'$, where it will remain again with overwhelming 
probability until the next transition. 

\section{Discussion and Path Forward}
\label{sec:discusssion}

We have extended the concept of currents and fluxes,
generated in non-equilibrium (driven) stochastic processes to higher
dimensions. This has been done for both the continuous \cite{CCK16b} 
and discrete cases. In the
continuous case the higher-dimensional current characterizes the same process as
the standard stochastic currents, i.e., Langevin stochastic dynamics on a
manifold $X$ (of arbitrary dimension ${\rm dim}(X) = m$) with inhomogeneous
noise.

The results on discrete stochastic models, associated with CW complexes that
are presented in this manuscript, focus on the periodic driving case with
emphasis on the adiabatic driving limit, and special emphasis on the effect of
quantization of stochastic currents in the low-temperature adiabatic limit, the
effect that in the case of graphs has been observed 
experimentally~\cite{LWDZ03, S09, AA07, P98} and
established theoretically~\cite{CKS12a, CKS13} in our earlier work.

We cannot overly emphasize that, as opposed to the continuous case \cite{CCK16b}, the discrete setting for stochastic fluxes
brings in a new class of discrete stochastic models, which are associated with
CW complexes, rather than just graphs.
The models associated with CW complexes apparently look like
a very artificial and unnatural generalization of Markov chain processes on
graphs. On the other hand, the higher-dimensional fluxes, despite needing a
variety of mathematical techniques (that are not very traditional in the
chemical physics community) to be properly handled, are conceptually very
simple. One deals with the same class of Langevin processes, with the only
difference that the new observables are associated with stochastic trajectories
of extended objects, namely higher-dimensional cycles. Alternatively,
discrete models deal with finite-dimensional vector spaces of states and thus
are advantageous computationally. This is one of the reasons why we spent so
much effort and a variety of techniques on considering the continuous models
(sections~(2) and (3) in the prequel \cite{CCK16b}) so
that in section~\ref{sec:Markov-chain} we were able to derive the CW complex
models, interpreting them as a computational tool for treating
higher-dimensional currents/fluxes in continuous spaces.

The main result presented in this paper, namely, rational quantization of higher dimensional currents in the low-temperature adiabatic limit, has been achieved by applying the following key steps.

(i) We started with deriving an expression for the average flux in the
adiabatic limit that generalizes the expression, obtained in our earlier work
for Markov periodic processes on graphs, and can be also interpreted as the
discrete counterpart of Eq.~(36) of \cite{CCK16b}, the 
latter describing the
flux in the continuous case. The derivation is straightforward and follows the
lines of the similar derivation presented in the first 
manuscript.

(ii) In the graph case the Boltzmann distribution is known explicitly, whereas
the pseudo-inverse that solves the Kirchhoff problem is expressed as a weighted
sum over the spanning trees of the graph. In the CW complex case it was not
obvious what an adequate definition of a spanning tree was, since various
definitions have been presented in the literature. Besides,
identifying an explicit formula for higher-dimensional Boltzmann distributions
appeared to be a much less tractable problem. We have made a surprising
observation that the Kirchhoff problem and the problem of identification of the
Boltzmann distribution are in fact dual to each other, and can be both treated
as finding a proper pseudo-inverse. We found a theorem in linear algebra, which
seems not to be a part of common knowledge, applied it to both problems, which
allowed to identify the adequate notions for spanning trees and co-trees, the
latter being objects naturally appearing in representing the pseudo-inverse for
the Boltzmann problem.

(iii) We further expressed the coefficients in front of the spanning trees and
co-trees in the aforementioned weighted sums over the trees and co-trees,
respectively, in terms of topological invariants of the CW complex and its
trees and co-trees, which resulted in a formula, expressed in completely
topological terms.

(iv) Finally we applied the low-temperature limit, resulting in a purely
topological formula that generalizes the expression for Markov chain processes
on graphs to the CW complex case. It needs to be stressed that as opposed to
integer quantization in the graph case, in a generic situation quantization is
rational for CW complex models. Formally it occurs due to additional factors in
denominators of the coefficients in front of the spanning trees and co-trees in
the higher-dimensional Kirchhoff tree and co-tree theorems, that are due to the
torsion (finite cyclic) subgroups of the relevant homology groups of the
CW complex, and its relevant spanning tress and co-trees. 
%The physics reason
%for that will be briefly discussed later in this section.

The additional complexity of rational quantization  can be addressed starting 
from the discrete CW-models. The nontrivial
nature of the higher-dimensional case shows up in discrete model in a less
nasty way. We still have a closed equation for the probability distribution,
which can be efficiently twisted, however the closed equation involves the
infinite-dimensional Hilbert space of states, associated with the infinite
graph, introduced in section~\ref{sec:Markov-chain}. On the other hand, in many
cases dynamics that involves Hilbert spaces can be efficiently studied, using
e.g., spectral methods. In either case, we see that that dimension of the
stochastic trajectory is responsible for the additional complexity and
not the underlying configuration space.

\section*{Acknowledgements}
This material is based upon work supported by the National Science Foundation under Grant No. CHE-1111350.

%\bibliography{higher-currents}{}
%\bibliographystyle{plain}

\end{document}